\documentclass[a4paper]{PoS}

\usepackage[utf8]{inputenc}
\usepackage{amsmath}
\usepackage[sort, compress, numbers]{natbib}

\title{Effects of magnetic fields on quark-antiquark interactions}

\ShortTitle{Effects of magnetic fields on quark-antiquark interactions}

\author{Claudio Bonati\footnote{Present address: Dipartimento di Fisica e Astronomia \& INFN Sez. Firenze, Via
Sansone 1, 50019, Sesto Fiorentino (FI), Italy}, Massimo D'Elia, Marco Mariti, Michele Mesiti, Francesco Negro, \speaker{Andrea Rucci}\\%
  Department of Physics of University of Pisa and INFN Pisa\\
  E-mail:  \email{bonati@df.unipi.it},
  \email{delia@df.unipi.it},
  \email{mariti@df.unipi.it},
  \email{mesiti@pi.infn.it},
  \email{fnegro@pi.infn.it},
  \email{andrea.rucci@pi.infn.it}
  }
\author{Francesco Sanfilippo\\
  School of Physics and Astronomy of University of Southampton, United Kindgdom\\
  E-mail: \email{fr.sanfilippo@gmail.com}}

\abstract{We discuss some recent results obtained in the study of
  strong quark-antiquark interactions in the presence of intense
  external magnetic fields by means of lattice QCD simulations. We
  confirm previous findings and show that both at zero and finite
  temperature the external field induces anisotropies in the static
  quark potential. An in-depth study suggests that the effects are
  essentially due to the variation of the string tension whose angular
  dependence can be nicely parametrized by the first allowed term in a
  Fourier expansion. In the confined phase at high temperature, we
  observe that the suppression of the string tension is enhanced as
  the strength of the external field increases. Our results support
  the idea that the loss of confining properties is the dominant
  effect related to the decrease of $T_c$ as a function of $B$.}

\FullConference{34th annual International Symposium on Lattice Field Theory\\
                 24-30 July 2016\\
                 University of Southampton, UK}

\begin{document}

\section{Introduction}
In many situations, such as heavy ion collisions~\cite{Skokov:2009qp}
or in the early Universe~\cite{Vachaspati:1991nm}, large magnetic
fields with intensities of $|e|B\gtrsim m_\pi^2$ are produced and
affect the properties of strongly interacting matter. In these
contexts, the magnetic fields may lead to significant modifications of the
properties of QCD. Indeed, even if gluons are not directly coupled to
electro-magnetic fields, substantial contributions can arise from
quark loops. This is confirmed both by analytical studies that use
effective field theories or perturbation theory and by many evidences
from lattice investigations (see, e.g., Refs.~\cite{Kharzeev:2012ph,
  Miransky:2015ava} for general reviews).

Important modifications may be expected on the static potential
between heavy quarks. In the explorative study in
Ref.~\cite{Bonati:2014ksa} it has been found that both the Coulomb
term and the linear confining piece of the potential become
anisotropic when a strong external magnetic field is turned on. This
has also been predicted by several model studies and could have
phenomenologically relevant effects in the heavy meson spectrum and
production in heavy ion collision~\cite{Hattori:2015aki, Guo:2015nsa,
  Fukushima:2015wck, Gubler:2015qok, Bonati:2015dka}.  In our study we
carry on the investigation of these effects on the potential at zero
temperature and explore also the confined phase in the finite
temperature regime. At $T=0$, a quantitative description of the
behaviour of the string tension and of its angular dependence is
provided, together with hints about its fate in the large field
limit. Then we move to the exploration of the $T>0$ regime, in order
to check whether the anisotropic effect survives, and also to
investigate the role of the magnetic field in the string tension
suppression near the deconfinement transition.

\section{Numerical setup}
In our work we consider QCD with $N_f=2+1$ quark flavours discretized
using a Symanzik tree-level improved gauge action and stout improved
rooted staggered fermions. The partition function reads
\begin{equation}\label{partfunc}
  Z(B) = \int \!\mathcal{D}U \,e^{-\mathcal{S}_{Y\!M}}
  \!\!\!\!\prod_{f=u,\,d,\,s} \!\!\!
  \det{({D^{f}_{\textnormal{st}}[B]})^{1/4}}\ .
\end{equation}
where $\mathcal{D}U$ is the measure over the gauge link variables,
$\mathcal{S}_{Y\!M}$ is the gauge action~\cite{weisz, curci} involving
the real part of the traces over $1\times1$ and $1\times2$ loops and
$D^{f}_{\textnormal{st}}[B]$ is the staggered fermion matrix with
two-times stout-smeared gauge links~\cite{morning}.

In order to consider the effect of an external background magnetic
field, in the continuum we introduce the abelian gauge four-potential
$a_\mu$ into the quark covariant derivative which takes the form
$D_\mu=\partial_\mu+igA^a_\mu T^a+iq_fa_\mu$ with $q_f$ electric
charge related to the quark flavour $f$. In the discretized theory,
where this operator is written in terms of the SU(3) gauge links
$U_{i:\mu}$ (with $i$ lattice site and $\mu$ direction), the insertion
of the electromagnetic field corresponds to the substitution
$U_{i:\mu}\to u_{i:\mu}U_{i:\mu}$ where $u_{i:\mu}$ is a U(1) phase
related to $a_{\mu}$. We consider external fields, hence $a_{\mu}$
is non-propagating: no $a_{\mu}$ kinetic term is added to the
Lagrangian.

In our case we work with a constant and uniform magnetic
field~\cite{Kharzeev:2012ph}. If we choose to fix $\vec{B}$ along one
of the lattice axes (say $\hat{z}$) then it is possible to show that,
due to the lattice periodic boundary conditions, the quantization
condition $eB_z={6 \pi b_z}/{(L_xL_y)}$ with $b_z\in\mathbb{Z}$ must
hold, where $e$ is the electric charge unity and $L_x$ and $L_y$ are
the lattice extents along $\hat{x}$ and $\hat{y}$. If $\vec{B}$ is not
aligned along one of the axes, we can consider each component
separately and write the total abelian phase as the product of them
over the three axes~\cite{Bonati:2016kxj}. Each component of $\vec{B}$
must satisfy an independent quantization condition and $\vec{B}$ will
be associated to an integer vector $\vec{b}$. If $L_x=L_y=L_z$ the
magnetic field vector $\vec{B}$ is proportional to $\vec{b}$.

We performed simulations at physical quark masses, i.e. tuning the
parameters according to the line of constant physics reported in
Refs.~\cite{physline1, physline2}. Note that the presence of the
external field does not affect the lattice
spacing~\cite{Bali:2011qj}. At zero temperature we used four lattices
$24^4$, $32^4$, $40^4$ and $48^3\times96$ with spacings ranging from
$a=0.2173~\rm{fm}$ to $a=0.0989~\rm{fm}$, and for each of them, we
performed runs with magnetic field quanta up to $|\vec{b}|=40$; this
give us access to fields roughly up to $|e|B\sim1~\rm{GeV}^2$. For
$T>0$ we used lattices $48^3\times N_t$ with $N_t=\{20,16,14\}$ and
$a=0.0989~\rm{fm}$, corresponding to physical temperatures sligthly
below $T_c$. In all cases, gauge configurations have been sampled by
the Rational Hybrid Monte-Carlo algorithm~\cite{Clark:2006wp},
collecting statistics of the order of $10^3-10^4$ trajectories for
each value of $\vec{b}$.

At $T=0$, the static $Q\bar{Q}$ potential $V(r)$ is extracted from
planar Wilson loops. Note that loops oriented along different
directions cannot generically be averaged since the $O(3)$ rotational
symmetry is broken to $O(2)$ by the external magnetic field. At finite
$T$, instead, we extracted the free energy $F_{Q\bar{Q}}(r,T)$ of a
quark-antiquark pair from Polyakov loop correlators.

\section{Analysis and results}
We adopted the following strategy: first, we determined the values of
the potential parameter at $T=0$ and $|e|B=0$ in order to use them as
a reference for the subsequent analysis. Then we investigated the
influence of the external field, performing a continuum limit
extrapolation of all the relevant observables. Finally, we moved on to
the $T>0$ regime, by working on our finest lattice to investigate
whether the $T=0$ anisotropies survive.

\subsection{T=0}
It is known that the $Q\bar{Q}$ potential can be well described by the
Cornell parametrization~\cite{Eichten:1974af}
\begin{equation}\label{eq:cornell_model}
V(r)=-\frac{\alpha}{r}+\sigma r + V_0\ ,
\end{equation}
where $\alpha$ is the Coulomb term, $\sigma$ is the so-called string
tension and $V_0$ is a constant. This functional form fits very well
our data at $T=0$ and $B=0$ with parameters\footnote{The statistical
  error is obtained from the $\mathcal{O}(a^2)$ fit of our data (the
  largest lattice spacing has been discarded); the systematic
  uncertainty has been estimated by using also the $\mathcal{O}(a^4)$
  fit model.}  $\alpha=0.395(22)(26)$ and
$\sqrt{\sigma}=448(20)(09)~\rm{MeV}$, that will be used as reference
in the following, corresponding to a value of the Sommer
parameter~\cite{Sommer:1993ce} $r_0=0.489(20)(04)~\rm{fm}$.

Turning the magnetic field on, we expect the potential to become
anisotropic and to acquire different values in the directions
orthogonal or parallel to $\vec{B}$~\cite{Bonati:2014ksa}. In general,
one can promote the potential $V(r)$ to a function $V(r,\theta,B)$ of
the magnetic field and of the angle $\theta$ between $\vec{B}$ and the
$Q\bar{Q}$ orientation. This general parametrization takes into
account the residual cylindrical symmetry around $\vec{B}$; we will
also impose symmetry under the reflection $\vec{B}\to-\vec{B}$. Then,
assuming the potential to be of the Cornell form
\eqref{eq:cornell_model} for all values of $\theta$, the most general
form of $V$ compatible with these symmetries is
\begin{equation}\label{eq:cornell_angular}
V(r,\theta;B) = -\frac{\alpha(\theta;B)}{r}+\sigma(\theta;B)r+V_0(\theta;B) \qquad
\mathcal{O}(\theta;B)=\Big(1-\sum_{n\ge1}c^{\mathcal{O}}_{2n}(B)\cos(2n\theta)\Big)\bar{\mathcal{O}}
\end{equation}
where the potential parameters $\mathcal{O}=\{\alpha,\sigma,V_0\}$
bring all the dependence on $\vec{B}$.

\begin{figure}[t]
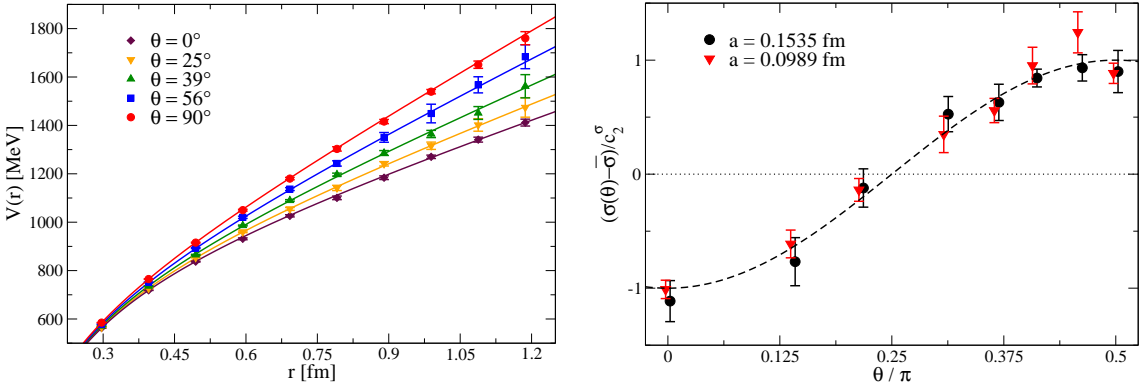

\includegraphics*[width=0.48\textwidth]{L48_potential_angles.eps}$\quad$
\includegraphics*[width=0.475\textwidth]{sigma_angular_dependence.eps}
\caption{\emph{Left}: the $Q\bar{Q}$ potential from the $48^3\times96$
  lattice for some values of $\theta$ at
  $|e|B\simeq1~\rm{GeV}^2$. \emph{Right}: angular variation of the
  string tension compared with the function $-\cos(2\theta)$ (dashed
  line).}
\label{fig:angular_B1GeV}
\end{figure}

In order to verify the reliability of this model, we analyzed the
angular dependence of the static potential on our two finest lattices
(corresponding to the lattice sizes $48^3\times96$ and $32^4$) for
three different orientations of the magnetic field with $|b|\simeq32$,
giving us access to 8 different $\theta$ angles. Results are shown in
Fig.~\ref{fig:angular_B1GeV}.
As expected from the preliminary data in Ref.~\cite{Bonati:2014ksa},
the potential $V(r)$ increases with the angle, and the model in
Eq.~\eqref{eq:cornell_angular} fits very well our data. We found that
the values $\bar{\mathcal{O}}$, which are a sort of average of the
parameters over the angles, are all compatible with those at
$B=0$. Morover, it turns out that it is sufficient to consider only
the first terms $c_2^{\mathcal{O}}$ in the Fourier series\footnote{If
  successive terms $c_{2n}$ with $n>1$ are included in the fit model,
  they turn out to be all compatible with zero.}, as show in
Fig.~\ref{fig:angular_B1GeV}.%  We also found that $V_0$ becomes
% anisotropic and that the values of the $c_2^{\mathcal{O}}$s decrease
% moving from the coarsest to the finest lattice
% spacing~\cite{Bonati:2016kxj}.

We performed a continuum limit extrapolation to determine whether
these anisotropies are only lattice artifacts or truly physical
effects. A complete determination of the angular distributions and of
the $\vec{B}$ dependence would require many simulations with different
lattice spacings and several values and orientations of the magnetic
field. However, since only the $c_2$s terms are needed in the
description of the angular dependence (at least with our precision),
we can simplify the task considerably by considering the two
quantities
\begin{equation}\label{eq:anisotropy}
\delta^{\mathcal{O}}(|e|B)=
\frac{\mathcal{O}_{XY}(|e|B)-\mathcal{O}_{Z}(|e|B)}{\mathcal{O}_{XY}(|e|B)+\mathcal{O}_{Z}(|e|B)}
\qquad R^{\mathcal{O}}(|e|B)=
\frac{\mathcal{O}_{XY}(|e|B)+\mathcal{O}_{Z}(|e|B)}{2 \mathcal{O} (|e|B = 0)}\ .
\end{equation}
These are, respectively, the anisotropy and the average value of the
parameter $\mathcal{O}$ taken along the $\hat{x},\hat{y}$ and
$\hat{z}$ directions. Using the fact that $c_{2n}\simeq0$ when $n>1$,
it is easy to show that
$\delta^{\mathcal{O}}(|e|B)\simeq c_2^{\mathcal{O}}(|e|B)$ and
$R^{\mathcal{O}}(|e|B)\simeq\bar{\mathcal{O}}(|e|B)/\mathcal{O}(|e|B=0)$. Therefore,
we can access all the information about the anisotropies just by
measuring the potential in the directions parallel or orthogonal to
the magnetic field.

\begin{figure}[t]
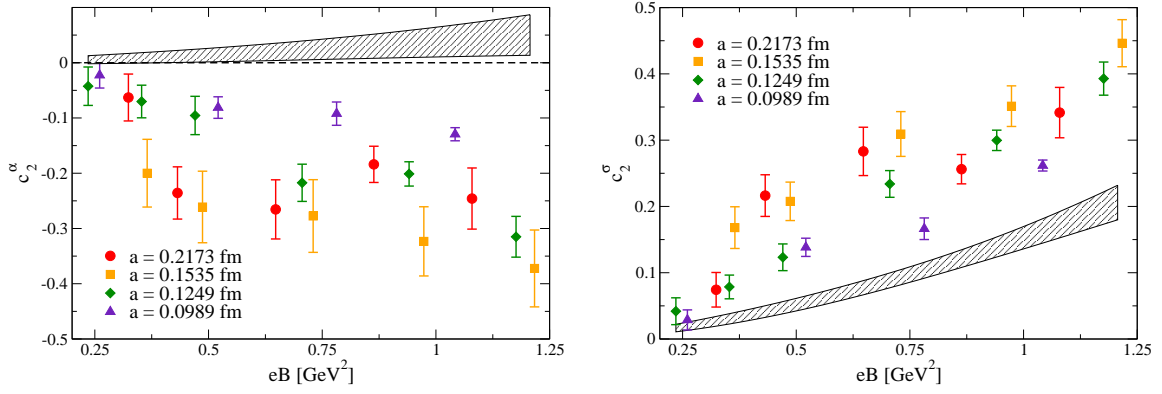

\centering
\includegraphics*[width=0.484\textwidth]{alpha_aniso.eps}$\quad$
\includegraphics*[width=0.479\textwidth]{sigma_aniso.eps}
\caption{The coefficients $c^{\mathcal{O}}_2$ of $\alpha$
  (\emph{left}) and $\sigma$ (\emph{right}). Bands are obtained
  through a continuum limit extrapolation on the data of our three
  finest lattices.}
\label{fig:anisotropy}
\end{figure}

Results for the anisotropies of $\alpha$ and $\sigma$, together with
their continuum limit extrapolations, are shown in
Fig.~\ref{fig:anisotropy}.
The fit procedure has been performed using a general power law for the
coefficients as a function of $|e|B$, with $\mathcal{O}(a^2)$
corrections, and it gave us very good values of the $\chi^2$
test~\cite{Bonati:2016kxj}. In the case of the string tension, the
anisotropy $c_2^{\sigma}$ survives in the continuum limit and grows
with the magnetic field. Conversely, the continuum limit of
$c_2^{\alpha}$ is compatible with zero, like for
$c_2^{V_0}$~\cite{Bonati:2016kxj}. We can therefore conclude that the
effect of the magnetic field on the static potential persists in the
continuum limit and is mostly due to the variation of the string
tension.

\begin{figure}[b]
\centering
\includegraphics*[width=0.48\textwidth]{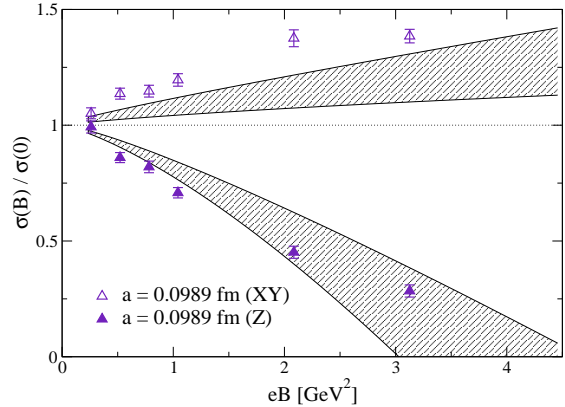}
\caption{String tension as a function $|e|B$. Bands are obtained
  through a continuum limit extrapolation performed only in the range
  $|e|B\lesssim1~\rm{GeV}^2$.}
  \label{fig:largeB}
\end{figure}

Finally, our findings suggest a sort of anisotropic deconfinement at
very large magnetic fields (see Fig.~\ref{fig:largeB}).
Indeed, for $|e|B\sim4~\rm{GeV}^2$ the string tension is predicted to
vanish in the longitudinal direction by extrapolating our fit
results. Data obtained for large $|e|B$ seem to roughly agree with
this behaviour, however a precise continuum extrapolation is difficult
in this region due to large cut-off effects expected at
$|e|B\sim a^{-2}$. By now, the presence of this sort of critical value
of $|e|B$ is to be taken as a fascinating speculation.

\subsection{T>0}
\begin{figure}[t]
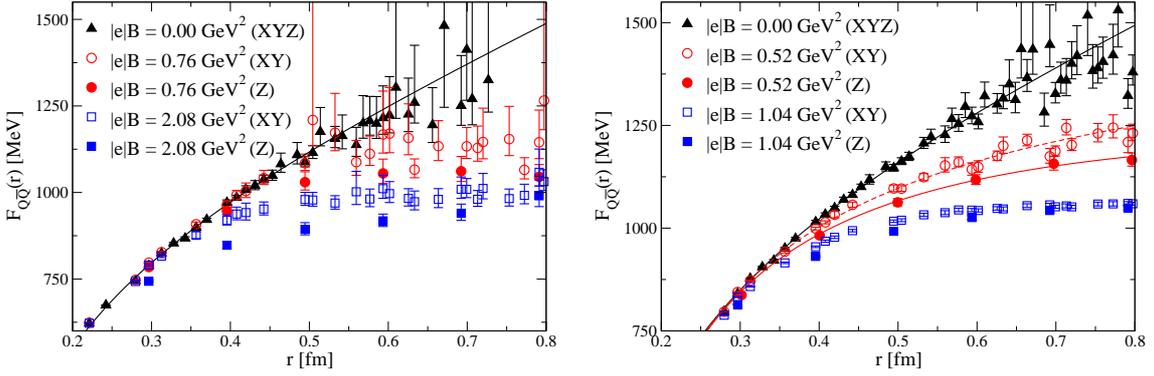

\includegraphics*[width=0.48\textwidth]{corr_finiteT_nt20.eps}$\quad$
\includegraphics*[width=0.48\textwidth]{corr_finiteT.eps}
\caption{Free energy $F_{Q\bar{Q}}(r,T)$ from $48^3$ lattices for two
  different temperature $T\simeq100~\rm{MeV}$ (\emph{left}) and
  $T\simeq125~\rm{MeV}$ (\emph{right}). Curves are obtained fitting
  the Cornell form and are shown only in the cases in which this
  parametrization reveals to be a good description of the data.}
\label{fig:pot_finiteT}
\end{figure}

\begin{figure}[b]
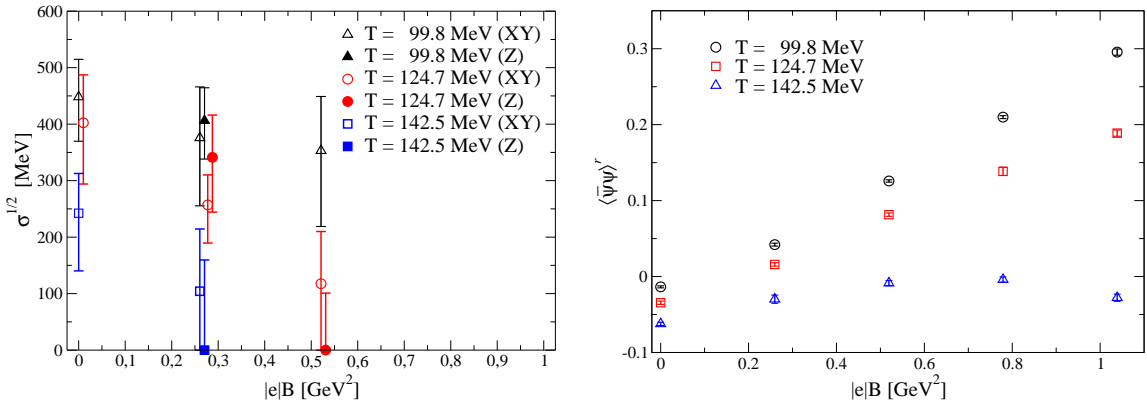

\includegraphics*[width=0.48\textwidth]{sigma_finiteT_vs_eB.eps}$\quad$
\includegraphics*[width=0.48\textwidth]{condensate.eps}
\caption{Square root $\sqrt{\sigma}$ of the string tension
  (\emph{left}) and the renormalized chiral condensate~\cite{Bali:2011qj} (\emph{right})
  as function of the magnetic field for some temperatures below
  $T_c$.}
\label{fig:string_and_chiral_condensate_vs_eB}
\end{figure}

Results for the free energy of a static $Q\bar{Q}$ pair at
temperatures lower than $T_c\sim155~\rm{MeV}$ are shown in
Fig.~\ref{fig:pot_finiteT}.
As one can see, the potential roughly follows the same behaviour
observed in the $T=0$ case, with the anisotropy still present. In this
temperature regime, the dominant effect turns out to be the flattening
of the potential as $B$ is increased. As a result, the Cornell form in
Eq.~\eqref{eq:cornell_model} ceases being suitable for the description
of the potential when both the temperature and the magnetic field
grow. This can be also seen from the behaviour of string tension
(leftmost plot in Fig.~\ref{fig:string_and_chiral_condensate_vs_eB}):
as expected, $\sigma$ tends to vanish as $T$ reaches the pseudocritical
temperature $T_c$, but this effect is remarkably enhanced by the
presence of the magnetic field.

These results are consistent with the picture of a decreasing chiral
pseudocritical temperature shown in previous
studies~\cite{Bali:2011qj,Bali:2012zg}. In the so-called inverse
magnetic catalysis, the decrease of $T_c$ is responsable of the
non-monotonic behaviour of the chiral condensate near the transition
temperature. Our data gives us evidences of the same phenomenon at the
level of the confining properties. Morover, the latter seems to be
dominant effect: in the range of $B$ and $T$ explored (see right plot
Fig.~\ref{fig:string_and_chiral_condensate_vs_eB}), the inverse
magnetic catalysis is hardly noticeable, while the flattening of the
potential is clearly visible.

\section{Conclusions}
It has been confirmed that the static $Q\bar{Q}$ potential is strongly
affected by an external magnetic field, becoming anisotropic. Studying
its angular dependence and performing continuum extrapolations at
$T=0$ we have shown that the anisotropy is a physical effect that
persists in the continuum limit. This is mostly due to the variation
of the string tension, whose angular dependence is very well described
by just the first Fourier term $c_2$. In the high-temperature regime,
the anisotropy is reduced and the main effect of the magnetic field is
the precocious disappeareance of the confining properties, indicated
by the flattening of the potential at large distances and by the
vanishing of the string tension, something that can be called
\emph{deconfinement catalysis}. This phenomenon can be understood in
terms of a decreasing $T_c$ and is noticeable even at low values of
the magnetic field, when there are no signals of inverse magnetic
catalysis in the chiral condensate.

\acknowledgments We acknowledge PRACE for giving us access to FERMI
machine at CINECA in Italy, under project Pra09-2400 - SISMAF.  FS
received funding from the European Research Council of the European
Community Seventh Framework Programme (FP7/2007-2013) ERC grant
agreement No 279757.  FN acknowledges financial support from the INFN
project SUMA.

\bibliographystyle{unsrt}
\setlength{\bibsep}{0pt plus 0.3ex}

\end{document}